\documentclass[12pt]{article}

\usepackage{arxiv}

\usepackage{setspace}
\usepackage[round,longnamesfirst]{natbib}
\usepackage{verbatim}
\usepackage{color}
\usepackage[pdftex,pdfborder={0 0 0 0}]{hyperref}
\usepackage[table]{xcolor}
\usepackage{amsmath}	
\usepackage{array,multirow}
\usepackage{lscape}
\usepackage{booktabs}
\usepackage{float}
\usepackage[caption = false]{subfig}
\usepackage[demo]{graphicx}
\usepackage{charter}
\usepackage[toc,header,title, page]{appendix}
\usepackage{titletoc}
\usepackage{enumitem}
\usepackage{titlesec}

\begin{document} 

\title{Uncertain Neighbors: Bayesian \\ Propensity Score Matching for Causal Inference}

\author{R. Michael Alvarez \\
        Professor of Political and Computational Social Science\\
        California Institute of Technology\\
        \texttt{\url{rma@caltech.edu}}\\
        \And
        Ines Levin \\
        Associate Professor of Political Science\\
        University of California, Irvine\\
        \texttt{\url{inelevin@uci.edu}}
        }

\date{\today}

\maketitle

\begin{abstract}We compare the performance of standard nearest-neighbor propensity score matching with that of an analogous Bayesian propensity score matching procedure. We show that the Bayesian approach makes better use of available information, as it makes less arbitrary decisions about which observations to drop and which ones to keep in the matched sample. We conduct a simulation study to evaluate the performance of standard and Bayesian nearest-neighbor matching when matching is done with and without replacement. We then use both methods to replicate a recent study about the impact of land reform on guerrilla activity in Colombia.
\end{abstract}

\section{Introduction}

\noindent Matching methods are often used by social scientists to estimate causal effects based on observational data or to adjust for randomization failures in the context of experimental or quasi-experimental research \citep{cr73,r79}. These methods allow comparing values of an outcome variable across observations that are similar in (observed) relevant ways except for differences in exposure to a presumed cause. A variety of procedures have been developed to determine the resemblance of observations and to match them conditional on similarity---often, conditional on an estimated measure of the distance between observations. One such distance measure is the propensity score, understood as the probability of exposure to a causal state or \emph{treatment}, conditional on observed correlates of treatment assignment \citep{hit97,rr83}. The true propensity score is an unobserved quantity, and is usually estimated using a regression approach.  Here we propose and evaluate the performance of a simple method for accounting for estimation uncertainty in the propensity score.

The idea of using estimated propensity scores for constructing matched samples emerged decades ago \citep{rr83,rr85}. While matching procedures continue to be developed (see for instance, \citealt{ds13,ikp12}), propensity score matching methods remain popular in applied work and researchers have kept on working on extensions and generalizations of propensity score-based procedures \citep{id04,ir12}.  The propensity score approach was developed as a way to address the dimensionality problems that researchers would face if they tried to control for observed differences between treatment and control observations by stratifying on multiple covariates. Estimated propensity scores can be used to match observations exposed to different causal states, using non-parametric matching procedures, in order to construct matched samples that are well-balanced on relevant covariates \citep{dw02}.  If the matching method succeeds in balancing all relevant confounders, then differences in outcomes across treatments can be attributed to differences in exposure to the presumed cause. 

\begin{figure}[p]
\caption{Estimated Propensity Scores}
	\centering
		\includegraphics[height=1.1\textwidth]{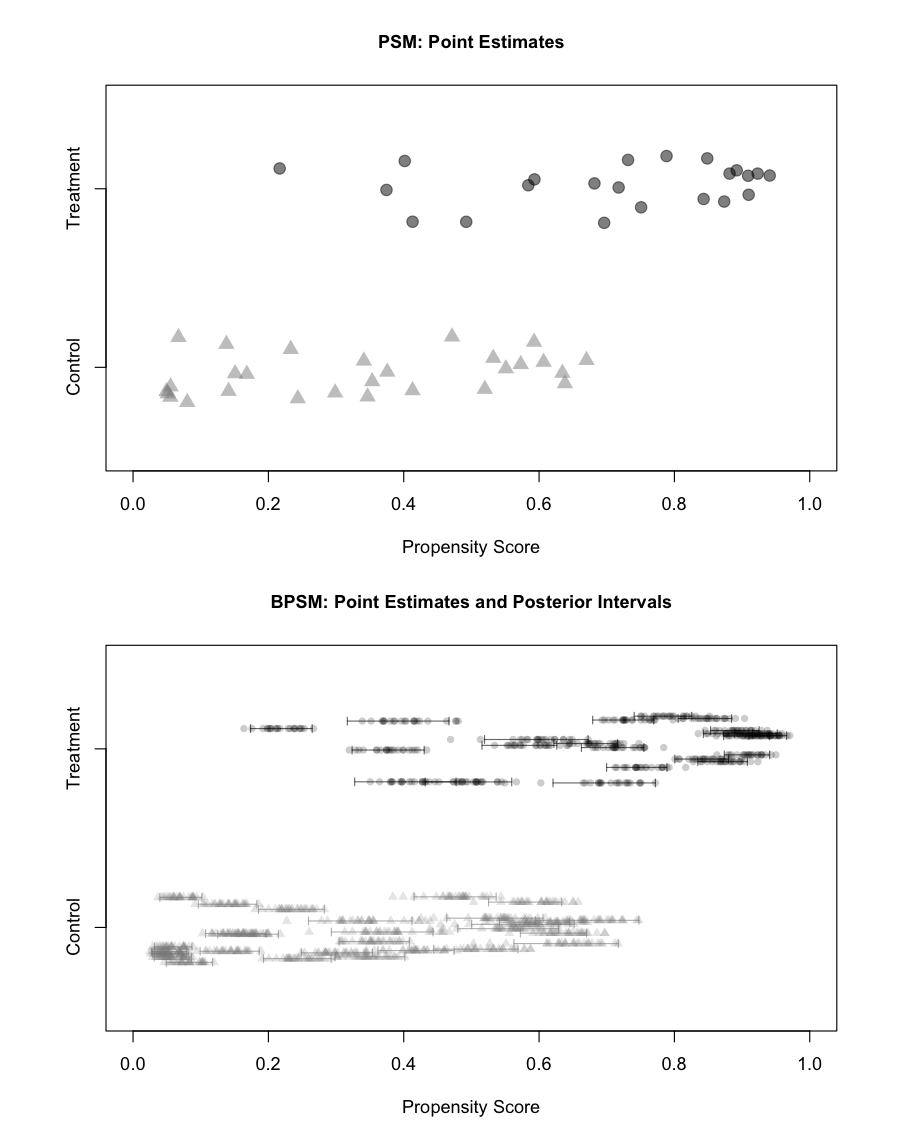}
 \centering{\parbox{5in}{\footnotesize{\vspace{0.5cm} Note: The figure illustrates information used (top-most plot) or ignored (bottom-most plot) by standard propensity score matching procedures, for 50 observations drawn at random from a synthetic dataset. While PSM only considers point estimates of propensity scores, the Bayesian propensity score matching procedure proposed in this paper takes into account information about other aspects of the posterior distribution of estimated propensity scores, including  estimation uncertainty.}}}\
	\label{fig:jitter}	
\end{figure}

Despite the ease of use and popularity of the propensity score, it has a number of limitations. Since we do not know the true mechanism underlying treatment assignment, estimated propensity scores may deviate in unknown ways from true assignment probabilities, producing biased measures of causal effects \citep{r99}. Furthermore, propensity score matching procedures are usually conducted in two stages: one involving the estimation of propensity scores, and another one in which treated and control observations are matched based on point estimates of the distance measure \citep{s10}. This ignores the fact that propensity scores are themselves estimated quantities, and as such there is always some degree of measurement uncertainty associated with those estimates \citep{tu02}. This idea is illustrated in Figure \ref{fig:jitter} (constructed with synthetic data), where the upper plot depicts the distribution of point estimates of propensity scores among treatment and control units (information typically used as input to propensity score matching algorithms) and the lower plot depicts 95\% credible intervals for the propensity score corresponding to each observation in the sample (information typically disregarded by propensity score matching algorithms). Although analysts always have some uncertainty about estimated propensity scores, standard matching algorithms treat the distance measure as a fixed quantity that is known with certainty \citep{mccandless09}.

Recent studies have investigated the extent to which incorporating uncertainty in the propensity score would have an impact on standard errors associated with measurements of causal effects. \citet{tu02} sought to incorporate uncertainty in the propensity score using a bootstrap method, and found that doing so leads to larger standard errors associated with estimates of treatment effects. However, the reliability of the bootstrapping approach has been called into question \citep{abadie08}. \citet{mccandless09} were the first to propose using Markov Chain Monte Carlo (MCMC) methods to incorporate uncertainty in the propensity score. Consistently with the previous two studies, they found that incorporating uncertainty in the propensity score leads to wider Bayesian credible intervals in the context of propensity-score-based stratification and regression procedure. \citet{an10} followed a similar approach, using Bayesian methods to incorporate uncertainty in the distance measure into propensity score regression and matching procedures, but in contrast to \citet{mccandless09}, found that doing so leads to reduced uncertainty about treatment effects. 

Both \citet{mccandless09} and \citet{an10}'s Bayesian inference procedures involve the simultaneous estimation of propensity score and outcome models.\footnote{See also \citet{zd14}, who develop a Bayesian method for jointly estimating propensity score and outcome models that allows accounting for uncertainty in the specification of the propensity score model.} \citet{kaplan12} criticize the simultaneous-estimation approach on the basis that, by estimating both models simultaneously, outcome data are allowed to inform the estimation of the propensity score---a potentially problematic attribute of the procedure, since it may introduce selection bias. While the propensity score should incorporate information about the treatment-assignment mechanism, it should not incorporate information about the outcome or treatment effect. To address the previous issue, \citet{kaplan12} proposed first estimating the propensity score model using MCMC methods; then repeatedly estimating treatment effects (using regression or non-parametric approaches), each time considering a different sample from the posterior distribution of the propensity score; and finally computing the mean and variance of estimated treatment effects across samples. 

In this research note, we evaluate the performance of Bayesian propensity score matching (BPSM) using a sequential estimation procedure along the lines proposed by \citet{kaplan12}, which avoids the problems discussed in the previous paragraph while still allowing the incorporation of information about the uncertainty in the propensity score. We find that BPSM has important advantages relative to standard propensity score matching (PSM), including that---as is generally the case with Bayesian estimation procedures---it produces samples of model parameters that can be used to summarize results and compute quantities of interest such as measures of centrality and dispersion \citep{jackman00}, allowing the analyst to easily get a sense of the uncertainty about treatment affects. 

An important difference between this paper and previous studies of the performance of Bayesian propensity score regression and matching (such as \citealt{mccandless09}, \citealt{an10}, and \citealt{kaplan12}), is that we focus on procedures where control units that are not comparable to treatment units are dropped from the analysis and are not taken into account for the computation of treatment effects. Dropping control observations with no close matches in the treatment group is a standard practice that allows achieving better balance in the matched sample \citep{ho07}. We argue that accounting for estimation uncertainty is particularly important in the case of matching procedures that entail dropping observations, since the decision of whether to keep a unit is based on the estimated distance measure. Indeed, we find that another reason why BPSM improves upon standard PSM is that the decision of whether to keep or drop observations from the matched sample is done in a less arbitrary manner.

\begin{figure}[p]
\caption{Example of Drop/Keep Decision for Control Observations}
	\centering
		\includegraphics[height=0.9\textwidth]{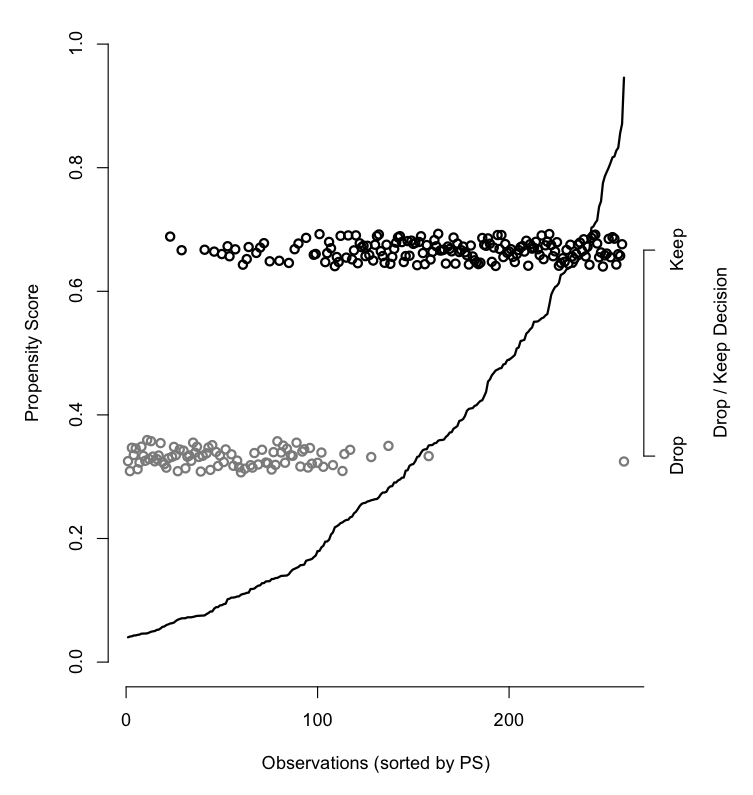}
  \centering{\parbox{5in}{\footnotesize{\vspace{0.5cm} Note: The figure depicts the relationship between point estimates of propensity scores and drop/keep decisions with a one-shot nearest-neighbor propensity score matching procedure, for a synthetic dataset containing 500 observations. Drop decisions are relatively rare for observations with point estimates of propensity score above a certain cutoff (around .2).}}}
	\label{fig:dropkeep}
\end{figure}

The arbitrariness of discarding units based on point estimates of propensity scores is illustrated in Figure \ref{fig:dropkeep}. This figure depicts the relationship between estimated propensity scores and the decision of whether to drop observations from the matched sample, when performing nearest-neighbor matching using calipers. In the case of control observations in the synthetic dataset used for constructing this example, dropping decisions are rare for point estimates of propensity scores above 0.4. This suggests that small changes in the distance measure (i.e. point estimates of propensity scores) may alter the composition of control units kept in the matched sample, especially for observations with point estimates of the propensity score close to the 0.4 cutoff. We argue that incorporating uncertainty in the propensity score by repeatedly matching treatment and control observations based on draws from the posterior distribution of propensity scores (instead of matching only once based on point estimates of the propensity score) can satisfactorily address the arbitrariness problem. By making more efficient use of the available information BPSM should result in more precise estimates of treatment effects, reflected in smaller standard errors and narrower confidence intervals.

It is questions like these that have led some prominent researchers to argue that propensity scores should not be used for matching (\citealt{king16}). \citealt{king16} argue that propensity score matching may actually increase imbalance, which of course is not what researchers want to achieve by using this approach for causal inference. We agree with this argument, that simple propensity score matching can be problematic.  Their insight is showing that propensity score matching seeks to achieve an outcome in the matched data that is something like a randomized experiment, rather than the more powerful design of the fully blocked experiment. They show that because of this issue,  propensity score matching drops observations from the analysis in an essentially random manner, which does not improve balance, and which leads to bias.  This is an important critique of propensity score matching, which we agree completely with.   However, we disagree with the advice provided by \citealt{king16}, because instead of
asserting that propensity score matching should not be used, we argue that merely by taking a Bayesian approach to estimation of the propensity score, we can resolve 
many of these problems with the propensity score.  That leads us to advocate the use of a Bayesian propensity score matching procedure, which is the central
contribution of our paper.  

\section{Methodology \label{sec:method}}

\noindent In this section we describe the propensity score matching procedures implemented in this paper. Let $Z_{i}$ denote an indicator of treatment assignment, which takes values 1 for individuals exposed or assigned to the treatment, and 0 for individuals in the control group. Further, let $Y_{zi}$ indicate the potential outcome for an individual $i$, which takes values $Y_{1i}$ if the individual is exposed to the treatment, and $Y_{0i}$ otherwise.

For each individual, $Y_{i}$ denotes the observed outcome and $X_{i}$ are observed covariates thought to affect both the outcome and the probability of exposure to the treatment, $P(Z_{i}=1|X_{i})$. If the outcome variable is a binary indicator that takes value 1 with probability $P(Y_{i}=1|Z_{i},X_{i})$, then the treatment effect can be defined as the change in $P(Y_{i}=1|Z_{i},X_{i})$ caused by a change in $Z_{i}$:\footnote{If the outcome variable is continuous, then the treatment effect can be defined as the change in the conditional expectation of $Y_{i}$, $E[Y_{i}|Z_{i},X_{i}]$, caused by a change in $Z_{i}$. That is, $TE_{i}=E[Y_{i}|Z_{i}=1,X_{i}=x]-E[Y_{i}|Z_{i}=0,X_{i}=x]$.}

\begin{equation}
TE_{i}=P(Y_{i}=1|Z_{i}=1,X_{i}=x)-P(Y_{i}=1|Z_{i}=0,X_{i}=x)
\end{equation}

We consider two alternative approaches to estimating average treatment effects: standard nearest-neighbor propensity score matching (PSM) and Bayesian nearest-neighbor propensity score matching (BPSM). In the latter case, we use MCMC methods to estimate the propensity score model, and then estimate the average treatment effects using a non-parametric propensity-score matching procedure. Next, we describe the main characteristics of the procedures that we use to measure average treatment effects, propensity score matching and Bayesian propensity score matching.

\subsection{Propensity Score Matching (PSM)} \label{desc_psm}

In the case of standard nearest-neighbor propensity score matching (PSM), we estimate the propensity score model using a logistic regression approach, such that:

\begin{equation} \label{psmodel}
logit[P(Z=1|X)]= X \Gamma
\end{equation}

\noindent where $\Gamma$ is a vector of coefficients and $X$ is a matrix of observed individual attributes.

Subsequently, we match treatment and control observations on the basis of a distance measure determined by the estimated propensity score, $\widehat{PS}=\hat{P}(Z=1|X)$. We search for matches using a non-parametric procedure, whereby treatment units are sorted in random order, and then each treatment unit is matched one at a time to the nearest control unit(s) along the distance measure.

We consider implementations of this procedure where the selection of control matches is alternatively done with and without replacement. Matching without replacement may lead to high bias, as some treated units may end up paired with `distant' control units. In anticipation of this issue, when performing matching without replacement we only consider control units within a maximum distance along the distance measure, selecting a single match at random from those control units lying within the chosen caliper.

\subsubsection{Average Treatment Effect}

To facilitate the interpretation of estimated treatment effects, we focus here on estimating the treatment effect for treated observations. This is achieved by focusing on matching procedures where all treated observations are kept in the matched sample and given equal weight in the estimation.

When the outcome variable is a binary indicator, $P_{0}$ denotes the estimated probability of success under the control; and $P_{1}$ denote the estimated probability of success under the treatment. For units within the matched sample, we compute $P_{0}$ as the average observed outcome taken over all matched control units:

\begin{equation}
P_{0} =  \frac{1}{\tilde{N}_{0}}\sum_{\hspace{0.25cm}i=1,i \in \tilde{C}}^{\tilde{N}_{0}} Y_{0i}
\end{equation}

\noindent where $\tilde{C}$ denotes the matched control group, $\tilde{N}_{0}$ denotes the number of matched control units, and $Y_{0i}$ denotes the observed outcome for individual $i$ in the matched control group. Similarly, we compute $P_{1}$ as the average observed outcome taken over all treatment units:

\begin{equation}\label{eqtreat}
P_{1} = \frac{1}{N_{1}}\sum_{\hspace{0.25cm}i=1,i \in T}^{N_{1}} Y_{1i}
\end{equation}

\noindent where $T$ denotes the original treatment group, $N_{1}$ denotes the number of treatment units, and $Y_{1i}$ denotes the observed outcome for individual $i$ in the treatment group. 

We compute the Average Treatment Effect on the Treated (ATT) as the difference between the probability of success under the treatment ($P_{1}$) and the probability of success under the matched control ($P_{0}$):
\begin{equation} \label{atepsm}
ATT = P_{1} - P_{0}
\end{equation}

By focusing on estimating the ATT, we make sure that estimates apply to a well-defined segment of the original sample (i.e. treated observations). We estimate the ATT by employing matching procedures where all treated observations are kept in the matched sample. Modifications of the matching procedure that help ensure better balance between the matched treatment and control units, such as matching each treatment unit to the nearest control unit within a certain maximum distance or \emph{caliper}, may lead to biased estimates of the ATT, as it may result in absence of suitable matches for some treatment units which are in consequence dropped from the matched sample.

While standard PSM produces point estimates of treatment effects, it does not produce readily available indicators of dispersion. Unless additional analyses are conducted after matching, it is not possible, with PSM, to evaluate the uncertainty of treatment effects.

\subsection{Bayesian Propensity Score Matching (BPSM)} \label{desc_bpsm}

First, we estimate a propensity score model similar to that given in equation (\ref{psmodel}), but using using MCMC methods. The Bayesian estimation procedure produces samples of the $\Gamma$ vector of parameters of the propensity score model, $(\hat{\Gamma}^{(1)} \ldots \hat{\Gamma}^{(K)})$, where $K$ denotes the total number of saved iterations. These samples can, in turn, be used to calculate samples of estimated propensity scores,  $(\hat{P}^{(1)} \ldots \widehat{PS}^{(K)})$, or samples of estimated linear predictors $(X \hat{\Gamma}^{(1)} \ldots X \hat{\Gamma}^{(K)})$, either of which can later be used as distance measures during the matching procedures.

For each $k$th sample, we match treatment and control units on the basis of the estimated propensity score, using a nearest-neighbor propensity score matching procedure similar to that described before. Subsequently, we compute the probability of success under the control for matched sample $k$, $P_{0}^{(k)}$, as the average observed outcome taken over all matched control units:

\begin{equation}
P_{0}^{(k)} =  \frac{1}{{\tilde{N}_{0}^{(k)}}}\sum_{\hspace{0.25cm} i=1,i \in \tilde{C}^{(k)}}^{\tilde{N}_{0}^{(k)}} Y_{0i}
\end{equation}

\noindent where $\tilde{C}^{(k)}$ denotes the matched control group in sample $k$ and $\tilde{N}_{0}^{(k)}$ denotes the number of matched control units in sample $k$. ust as for PSM, we focus on matching procedures where all treated observations are kept in the matched samples. The probability of success under the treatment ($P_{1}$) is thus calculated over all treated observations, just as in expression \ref{eqtreat}. For each matched sample $k$, we compute the average treatment effect as:

\begin{equation}
ATT^{(k)} = P_{1} - P_{0}^{(k)}
\end{equation}

Thus, the BPSM procedure does not produce a single point estimate of the ATT, but a sample of estimates of size $K$. This sample can be used to generate summaries of the posterior distribution of the ATT, including measures of centrality such as the mean, and measures of dispersion such as credible intervals.  Unlike PSM, BPSM can be used not only to estimate the magnitude of the ATT, but also to evaluate its associated uncertainty. 

\section{Simulation Studies}

In this section, we undertake a number of different simulation studies to examine the relative performance of the PSM and BPSM approaches.  We begin with the obvious simulation:  we know the specification of the propensity score and outcome models, and we estimate the true model using both approaches.  These simulations indicate that the BPSM approach is superior to PSM for recovering estimates of the treatment effects.  The second set of simulations considers a more likely scenario for applied researchers using observational data:  the propensity score model is misspecified due to the absence of a confounder.  Here we present simulation results from a variety of simple misspecifications, and in all cases we present evidence that the BPSM approach dominates the PSM approach.  Thus, the simulation studies presented here suggest that the BPSM approach is superior to the PSM approach.  

\subsection{Matching with Replacement}

We used Monte Carlo simulation to evaluate the performance of PSM and BPSM under a correctly specified propensity score and outcome model, for different sample sizes (500 and 2,000). For each combination of sample size and magnitude of treatment effects, we generated $J = 1,000$ synthetic data sets using the following procedure:

\begin{enumerate}
\item Generate J covariate matrices $(X^{(1)} \ldots X^{(J)})$, where each $X^{(j)}$ includes a vector of ones, a binary indicator $x_{1}^{(j)}$, and an order categorical variable $x_{2}^{(j)}$. At each step $j$ of the simulation procedure, draw variables $u_{1}^{(j)}$ and $u_{2}^{(j)}$ from a multivariate normal distribution, with $corr(u_{1}^{(j)}, u_{2}^{(j)}) = 0.25$. Variable $x_{1}^{(j)}$ is a binary indicator that equals one when a variable $u_{1}^{(j)}$ takes positive values and otherwise equals zero, and $x_{2}^{(j)}$ is a 6-point scale created by breaking $u_{2}^{(j)}$ into six intervals of equal length.

\item Conditional on simulated $X^{(j)}$'s and a fixed vector of parameters $\Gamma=[-6,2,1]$ (with the first element being an intercept, and the second and third elements capturing the impact of $x_1$ and $x_2$, respectively, on treatment assignment) generate J probabilities of treatment assignment (i.e. true propensity scores) $(PS^{(1)} \ldots PS^{(J)})$ based on a logistic regression model similar to that given in expression (\ref{psmodel}). Then, generate J indicators of treatment assignment $(Z^{(1)} \ldots Z^{(J)})$ by repeatedly drawing $Z^{(j)}$'s from a binomial distribution with probabilities given by true propensity scores $PS^{(j)}$'s.

\item Conditional on simulated $x_{1}^{(j)}$'s, $x_{2}^{(j)}$'s, $Z^{(j)}$'s, a fixed coefficient $\beta = 1$ capturing the effect of the treatment, and fixed parameters $\theta_{1} = 2$ and $\theta_{2} = -2$ capturing the impact of $x_1$ and $x_2$ on the outcome, respectively,  generate J probabilities of success $(P_{y}^{(1)} \ldots P_{y}^{(J)})$ based on a logistic regression model. Then, generate J binary outcome indicators $(Y^{(1)} \ldots Y^{(J)})$  by repeatedly drawing $Y^{(j)}$'s from a binomial distribution with probabilities given by simulated $P_{y}^{(j)}$'s.
\end{enumerate}

After generating $J = 1,000$ synthetic data sets, we applied PSM and BPSM procedures similar to those described in sections \ref{desc_psm} and \ref{desc_bpsm}. For each $j$th  data set, we discarded control units outside the support of the propensity score in the treatment group, and then performed one-to-one matching with replacement. The simulation procedure allowed us to evaluate both the bias and error associated with each matching procedure by comparing ATTs estimated using PSM and BPSM with true ATEs. True ATTs were computed by calculating individual treatment effects (which vary as a function of $x_{1i}$ and $x_{2i}$ for each individual, since the outcome model is non-linear) and then taking the average over the population, for each $j$ step of the simulation procedure. 

\begin{figure}
\caption{Simulation Study, Distribution of Bias}
  \includegraphics[height=\textwidth]{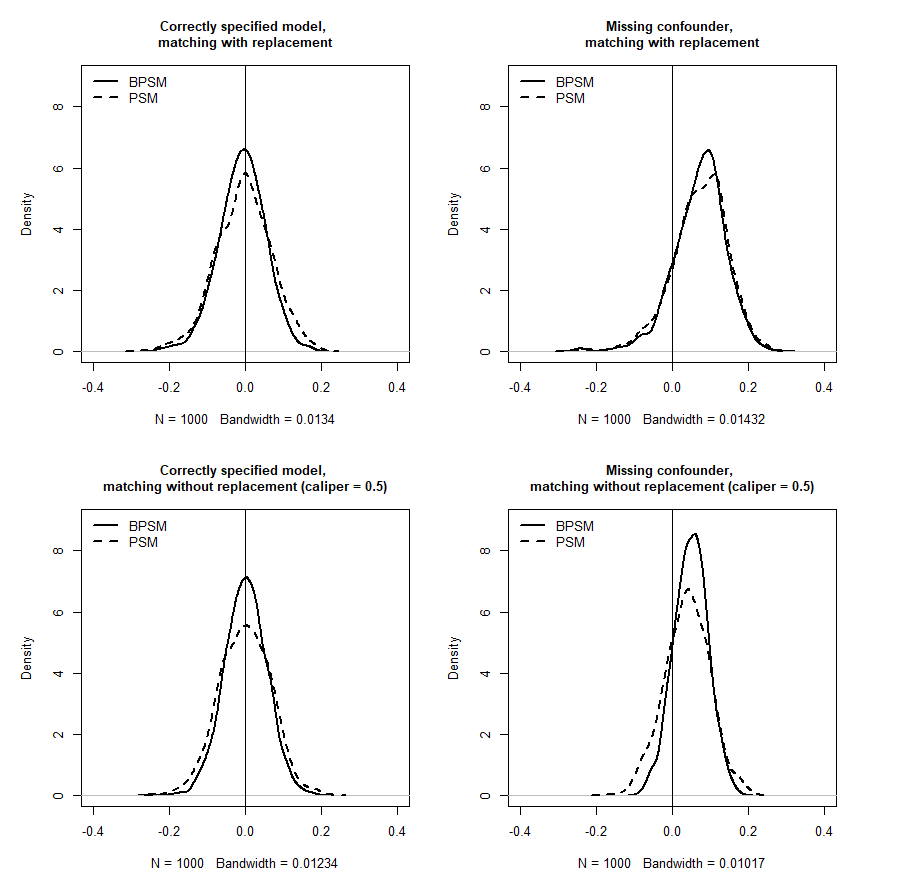}
  \centering{\parbox{5in}{\footnotesize{\vspace{0.5cm} Note: Results correspond to a simulation study conducted under the assumption of small sample size ($N = 500$) and small treatment effect ($\beta = .25$). Number of simulations: 1,000.}}}
\label{fig:distbias}
\end{figure}

The results of these simulations are shown in the upper-left plot of Figure \ref{fig:distbias} and in the upper section of Table A1 in the Online Appendix. With the configuration of parameters described above, about 43\% of observations are matched in the case of PSM, and similar percentages are matched on average at each iteration of BPSM. However, between 95\% (for N = 500) and 97\% (for N = 2,000) of observations are matched at least once (that is, are kept in the matched sample for at least one iteration of the matching procedure) for BPSM. These results suggest that treatment effects computed using BPSM incorporate information from a larger number of observations, compared to PSM. 

Both PSM and BPSM produces estimated ATTs that are relatively close to the truth. Differences in bias between procedures are not statistically significant at conventional confidence levels. BPSM estimates, however, are systematically less noisy compared to those produced by PSM. The spread and MSE of estimated ATTs are markedly lower for BPSM. By making less arbitrary decisions about which observations to keep or drop during the matching procedure, BPSM incorporates information from a larger number of observations, which results in higher accuracy.

\subsection{Additional Simulations}

We carried out similar simulations with the only difference that the propensity score model is misspecified due to the omission of a confounder.

The upper-right plot in Figure \ref{fig:distbias} corresponds to a situation where the missing confounder ($x_{3}$) is a 10-point scale generated based on a third variable $u_{3}$ drawn from the same multivariate distribution as $u_1$ and $u_2$ (the continuous variables used to determine $x_1$ and $x_2$), with $corr(u_{1}^{(j)}, u_{3}^{(j)}) = corr(u_{2}^{(j)}, u_{3}^{(j)}) = 0.25$, and a positive effect of $x_{3}$ on both $PS$ and $P_{y}$. Results suggest that the omission of a confounder leads to higher bias in estimated ATTs for both matching procedures (PSM \emph{and} BPSM). This results in higher MSE for both methods, but the amount of noise in the estimates is still lower for BPSM than for PSM (see lower section of Table A1 in the Online Appendix).

Lastly, we considered alternative implementations of both procedures where matching is done without replacement. In doing so, we searched for matches within a caliper of 0.5 standard deviations of the estimated propensity score (in the case of BPSM, this was repeated at each iteration). Results, shown in the bottom row of Figure \ref{fig:distbias}, are generally similar to those found with matching with replacement and without a caliper (also see Table A2 in the Online Appendix).

On the other hand, within-caliper nearest-neighbor matching requires dropping control units located far from the treatment group along the distance measure, as well as treatment observations that remain unmatched after all potential control matches have been used (this is the case because matching is conducted without replacement).  Thus, part of the bias can also be explained by the fact that the matched sample is not perfectly representative of the original sample, but biased toward specific values of observed covariates.

\section{Application: Land Reform and Insurgency in Colombia}

To demonstrate the practical utility of BPSM, we turn in this section to an application:  a replication of a study by \citet{albertus13} about the impact of land reform on guerrilla activity in Colombia. The central question investigated by \citet{albertus13} is whether land reform could serve as a tool for reducing guerrilla warfare, by addressing income inequality and improving the living conditions of peasants who might otherwise support insurgency. The authors investigate this question using municipality and municipal-year data covering a 12-year period ranging from 1988 to 2000. One of the methods that they employ to assess the impact of land reform is propensity score matching, where the treatment variable is a binary indicator of ``at least 300 plots reformed from 1988 to 2000, and at least three years with fifty or more plots reformed''  \citep[p. 215]{albertus13}. The outcome variable is a count of the number of guerrilla attacks recorded over the period.\footnote{Covariates included in the matching procedure include: prior plots reformed, paramilitary attacks, government attacks, poverty, population density, other tenancy, coca region, new colonized region, altitude, and percent minorities.} Contrary to expectations, the authors find that land reform is usually followed by an increase in the number of guerrilla attacks.

We do not take issue with the selection and measurement of variables entering the matching procedure, nor with the particular matching algorithm used by the authors (one-to-one propensity score matching with replacement, discarding units outside the common support of the propensity score). We merely observe that the use of one-to-one matching in a context where the number of control units greatly exceeds the number of treatment units, leads analysts to discard a large proportion of control units on the basis of the estimated propensity score.\footnote{In the case of the municipal-level data, the ratio of control to treatment units in the unmatched data set is larger than 13, and the use of one-to-one matching leads authors to drop 68\% of control units.} While most of the discarded control units are not comparable to units in the treatment group, some of them are borderline cases that are not significantly further away from the treatment than some of the matched control units. In situations like this, ignoring the uncertainty the propensity score might lead to arbitrariness in the selection of control units to be kept in the matched sample. In the rest of this section we show how results change when the Bayesian approach is used in order to account for estimation uncertainty in the propensity score.\footnote{In order to ensure that we used exactly the same data as \citep{albertus13}, we downloaded their replication package from \url{http://esoc.princeton.edu/files/land-reform-counterinsurgency-policy-case-colombia} (\citep{albertus13r}). We first replicated their analysis using the Stata code included in the replication package, as well as in R. Subsequently, we re-analyzed the data using BPSM.}

\begin{table}[h]
\caption{Replication of Albertus and Kaplan 2012, Table 3, Panel A}
  \centering
    \begin{tabular}{lccc}
 \multicolumn{4}{c}{a. Unit of analysis: Municipality} \\
     \toprule
    \toprule
& Units matched & ATE & SE \\
& at least once (\%) &  &  \\
       \midrule
PSM & 11.38 & 34.57 & 13.37 \\ 
 BPSM & 95.72 & 13.31 & 9.51 \\ 
    \bottomrule
 \\
 \multicolumn{4}{c}{b. Unit of analysis: Municipality-year} \\
     \toprule
& Units matched & ATE & SE \\
& at least once (\%) &  &  \\
       \midrule
 PSM & 10.38 & 1.40 & 0.54 \\ 
 BPSM & 60.34 & 1.04 & 0.27 \\ 
    \bottomrule
 \\
    \end{tabular}
      \centering{\parbox{5in}{\footnotesize{\vspace{0.5cm} Note: PSM indicates standard propensity score matching and BPSM indicates \\ Bayesian propensity score matching. For PSM, standard errors were computed using bootstrapping.  The implementation fo BPSM included post-matching regression adjustment.}}}\
  \label{tab:tab5lab}
\end{table}

\begin{figure}[p]
\caption{Land Reform and Insurgency in Colombia, Units Used During BPSM Procedure}
	\centering
		\includegraphics[width=1\textwidth]{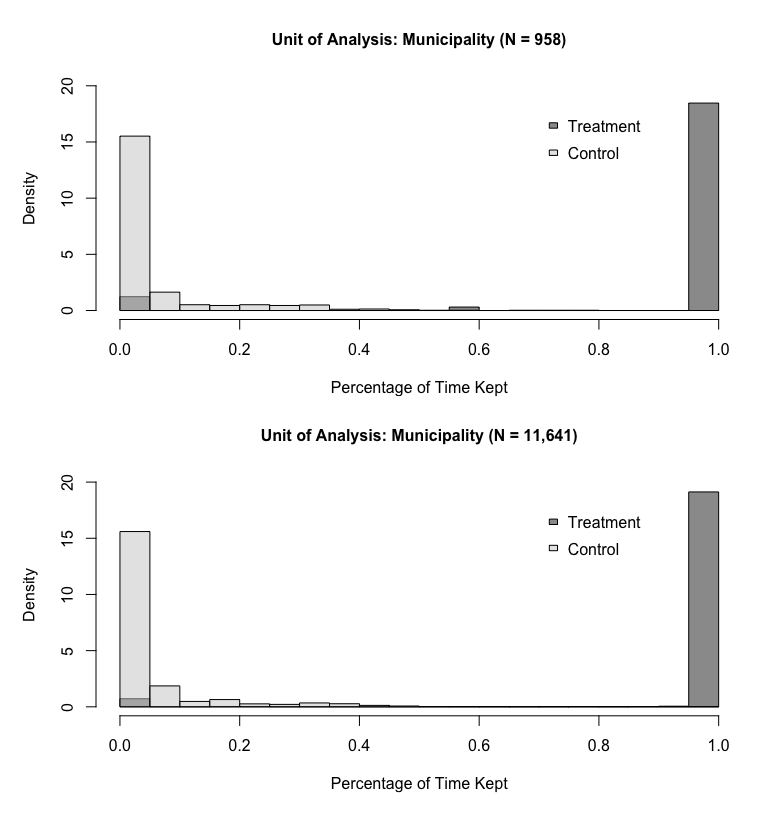}
	\label{fig:datause}
\end{figure}

Table \ref{tab:tab5lab} provides a comparison of the results found using a PSM procedure identical to the one used by \citet{albertus13}, and a BPSM procedure that is similar in every way. Panel (a) of the table gives results found using data at the municipality level, and panel (b) gives results found using data at the municipality-year level. Similarly to what we found in our simulation studies, the proportion of units kept in the matched sample after implementing PSM (11.38\% for the municipality data and 10.38\% for the municipality-year data) is considerably smaller than the proportion of units matched at least once during the implementation of the BPSM procedure (85.91\% for the municipality data and 49.76\% for the municipality-year data). However, the fact that some control units (which would otherwise be dropped under PSM) are used \emph{at least once} during the BPSM procedure, does not necessarily mean that they are used frequently. This feature is illustrated in Figure \ref{fig:datause}, which gives information about the proportion of the time that treatment units (dark grey bars) and control units (light gray bars) are kept in the matched sample during the BPSM procedure. While treatment units (a total of 65 in municipality dataset, and 754 in in the municipality-year dataset) are used almost all of time, most control units (a total of 893 in municipality dataset, and 10,887 in in the municipality-year dataset) are used less than 20\% of the time, with a majority being used rarely or never. Figure \ref{fig:datause} suggests that PSM and BPSM are actually not so different in terms of the amount of information that they incorporate; PSM also kept most treatment units and dropped most control units. Is the fact that some additional control units are matched a small proportion of the time under BPSM enough to affect estimates of treatment effects?

\begin{figure}[p]
\caption{Land Reform and Insurgency in Colombia, Average Treatment Effects}
	\centering
		\includegraphics[width=1\textwidth]{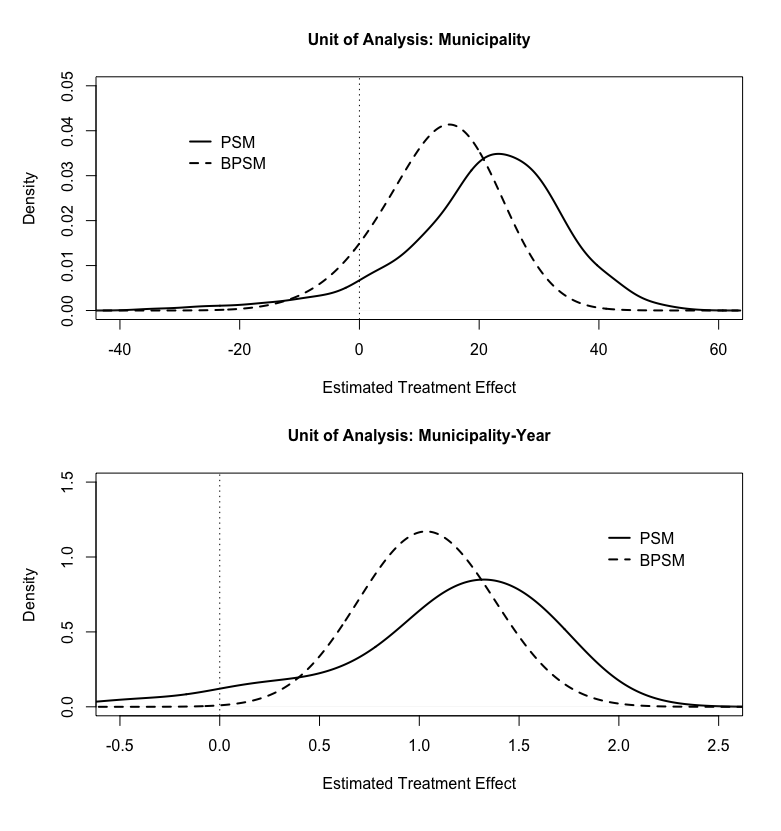}
	\label{fig:ate}
\end{figure}

Table \ref{tab:tab5lab} also gives information about the ATE found under each procedure. In their paper, \citet[Table 3, Panel A]{albertus13} report results corresponding to PSM, together with bootstrap estimates of standard errors. For the municipal-level data, they find that land reform leads to 34.57 more attacks, with a bootstrapped standard error of 11.68. When we replicate their analysis, we find similar results, although bootstrapped standard errors are slightly larger (13.37), which could happen for random reasons. When we re-analyze the data using BPSM (as described in Section \ref{sec:method}), however, we find that land reform leads to a positive but non-significant increase in the number of attacks (the ATE drops to 13.31, with standard error equal to 9.51, and 95\% credible interval ranging between -7.01 and 30.46). These findings can be visualized in the upper panel of Figure \ref{fig:ate}. This figure presents density curves and average values for the ATEs estimated using PSM and BPSM. The results obtained using BPSM are substantively similar to those found by \citet{albertus13} using PSM, though our BPSM ATEs are of smaller magnitude and higher variance than those that were presented in the original paper.  

We found comparable results for the municipality-year data. When matching is done based on point estimates of propensity scores, estimated ATEs are considerably larger than when matching is done repeatedly on the basis of draws from the posterior distribution of propensity scores. In their paper, \citet[Table 3, Panel A]{albertus13} report that land reform leads to 1.40 more attacks, with a bootstrap standard error of 0.61. When we replicate their analysis using PSM, we find similar results, though again the BPSM ATEs are of smaller magnitude than those estimated with propensity score matching.

\section{Conclusion}

Point estimates of propensity scores are commonly used for preprocessing the data by matching treatment and control units with similar predicted probabilities of assignment to treatment, in order to construct matched samples where covariates are well-balanced across treatment and control groups. The process typically involves dropping units outside the support of the propensity score; control units with very low ex-ante probability of being assigned to treatment; and sometimes treatment units for which no matches are available in the control group. While propensity score matching allows the analyst to address the dimensionality problem that would ensue if they tried to match units on the basis of multiple individual covariates, existing procedures also have a number of disadvantages. In particular, analysts tend to disregard the fact that estimated propensity scores have associated uncertainties. Standard approaches that take the propensity score as given can be problematic since they may cause the analyst to make arbitrary decisions regarding whether to keep or drop units from the matched sample.

We proposed a simple modification of standard propensity score matching procedures that can be easily implemented using Bayesian estimation. The Bayesian approach has several advantages, including that it can be use to calculate point estimates of treatment effects, as well as associated measures of uncertainty, without the need of resorting to bootstrapping or post-matching simulation procedures. Since matching under BPSM is done probabilistically instead of deterministically, it leads to less arbitrary decisions about whether to keep or drop observations from the matched sample. The results of our simulation study were in line with our expectations; they indicated that incorporating information about uncertainty in the propensity score leads to lower dispersion in estimates of treatment effects, compared to standard propensity score matching.   Furthermore, we replicated a published study that employed PSM and showed the utility of the BPSM approach in that application.  All of the evidence presented in this paper documents the utility of the BPSM approach.

\clearpage

\appendix

\setcounter{page}{1}

\appendixpage

\section*{\Large Supplementary materials for: ``Uncertain Neighbors: Bayesian \\ Propensity Score Matching for Causal Inference''}

\clearpage

\section*{Supplementary tables and figures}

\setcounter{table}{0}
\renewcommand{\thetable}{A\arabic{table}}

\setcounter{figure}{0}
\renewcommand{\thefigure}{A\arabic{figure}}

\section{Additional Tables and Figures}

\begin{table}[htbp]
\caption{Simulation Study, Matching with Replacement}
  \centering
    \begin{tabular}{lccccccc}

\multicolumn{8}{c}{a. Correctly specified propensity score model} \\
    \toprule
& Units matched &  & \multicolumn{2}{c}{95\% Interval} & & &  \\
& at least once (\%) & ATT &  2.5\% & 97.5\% & Bias & MAB & MSE  \\
      \midrule
 \multicolumn{8}{l}{\textit{N = 500}} \\
PSM & 42.6 & 12.7 & -2.2 & 26.9 & -0.4 & 5.8 & 7.3 \\ 
BPSM & 94.6 & 12.4 & 0.1 & 24.1 & -0.7 & 4.9 & 6.2 \\  
 \\
 \multicolumn{8}{l}{\textit{N = 2,000}} \\
PSM & 42.5 & 13.7 & 6.8 & 20.2 & -0.3 & 2.8 & 3.4 \\ 
BPSM & 96.8 & 13.5 & 7.7 & 18.9 & -0.5 & 2.3 & 2.9 \\ 
     \bottomrule
 \\
 \multicolumn{8}{c}{b. Incorrectly specified propensity score model} \\
  \multicolumn{8}{c}{(missing confounder)} \\
     \toprule
& Units matched &  & \multicolumn{2}{c}{95\% interval} & & &  \\
& at least once (\%) & ATT &  2.5\% & 97.5\% & Bias & MAB & MSE  \\
      \midrule
 \multicolumn{8}{l}{\textit{N = 500}} \\
PSM & 67.3 & 20.0 & 3.8 & 32.7 & 7.3 & 8.8 & 10.3 \\ 
BPSM & 98.5 & 20.0 & 4.7 & 32.2 & 7.3 & 8.5 & 10.0 \\ 
 \\
 \multicolumn{8}{l}{\textit{N = 2,000}} \\
PSM & 67.8 & 19.9 & 13.5 & 26.2 & 6.3 & 6.4 & 7.2 \\ 
BPSM & 99.3 & 19.9 & 13.9 & 26.1 & 6.3 & 6.4 & 7.0 \\ 
    \bottomrule
    \end{tabular}
    	\centering{\parbox{5in}{\footnotesize{\vspace{0.5cm} Note: Number of simulations: 1,000 for each combination of treatment effect size and sample size. The first column gives the percentage of observations kept in the matched sample (PSM) or retained \emph{at least once} during the matching procedure (BPSM). The next five columns provide the following summary measures for the estimated ATT: mean value and 95\% credible interval; bias (average difference between the true and estimated ATT); mean absolute bias or MAB (average of the absolute value of the difference between the true and estimated ATT); and mean squared error or MSE (average of the squared difference between the true and estimated ATT).}}}
  \label{tab:tab1lab}
\end{table}

\newpage

\begin{table}[htbp]
\caption{Simulation Study, Matching without Replacement, with Caliper}
  \centering
    \begin{tabular}{lccccccc}

\multicolumn{8}{c}{a. Correctly specified propensity score model} \\
    \toprule
& Units matched &  & \multicolumn{2}{c}{95\% Interval} &  & &  \\
& at least once (\%) & ATT &  2.5\% & 97.5\% & Bias & MAB & MSE  \\
      \midrule
 \multicolumn{8}{l}{\textit{N = 500}} \\
PSM & 40.4 & 13.1 & -0.9 & 25.7 & -0.0 & 5.4 & 6.7 \\ 
BPSM & 95.6 & 13.2 & 2.1 & 24.1 & 0.1 & 4.4 & 5.6 \\  
 \\
 \multicolumn{8}{l}{\textit{N = 2,000}} \\
PSM & 42.5 & 13.5 & 7.2 & 20.0 & -0.5 & 2.6 & 3.3 \\ 
BPSM & 98.6 & 14.0 & 8.8 & 19.7 & -0.0 & 2.2 & 2.8 \\
     \bottomrule
 \\
 \multicolumn{8}{c}{b. Incorrectly specified propensity score model} \\
  \multicolumn{8}{c}{(missing confounder)} \\
     \toprule
& Units matched &  & \multicolumn{2}{c}{95\% interval} & &  &  \\
& at least once (\%) & ATT &  2.5\% & 97.5\% & Bias & MAB & MSE  \\
      \midrule
 \multicolumn{8}{l}{\textit{N = 500}} \\
PSM & 46.6 & 16.6 & 4.7 & 28.2 & 3.9 & 5.8 & 7.1 \\ 
BPSM & 98.5 & 17.6 & 8.6 & 26.4 & 4.9 & 5.6 & 6.6 \\ 
 \\
 \multicolumn{8}{l}{\textit{N = 2,000}} \\
PSM & 47.8 & 16.2 & 10.2 & 22.5 & 2.6 & 3.4 & 4.1 \\ 
BPSM & 99.4 & 16.2 & 11.0 & 21.3 & 2.6 & 3.0 & 3.7 \\

    \bottomrule
    \end{tabular}
  		\centering{\parbox{5in}{\footnotesize{\vspace{0.5cm} Note: Number of simulations: 1,000 for each combination of treatment effect size and sample size. The first column gives the percentage of observations kept in the matched sample (PSM) or retained \emph{at least once} during the matching procedure (BPSM). The next five columns provide the following summary measures for the estimated ATT: mean value and 95\% credible interval; bias (average difference between the true and estimated ATT); mean absolute bias or MAB (average of the absolute value of the difference between the true and estimated ATT); and mean squared error or MSE (average of the squared difference between the true and estimated ATT).}}}
  \label{tab:tab2lab}
\end{table}

\clearpage

\begin{figure}
\caption{Simulation Study, Comparison of Bias}
  \includegraphics[height=\textwidth]{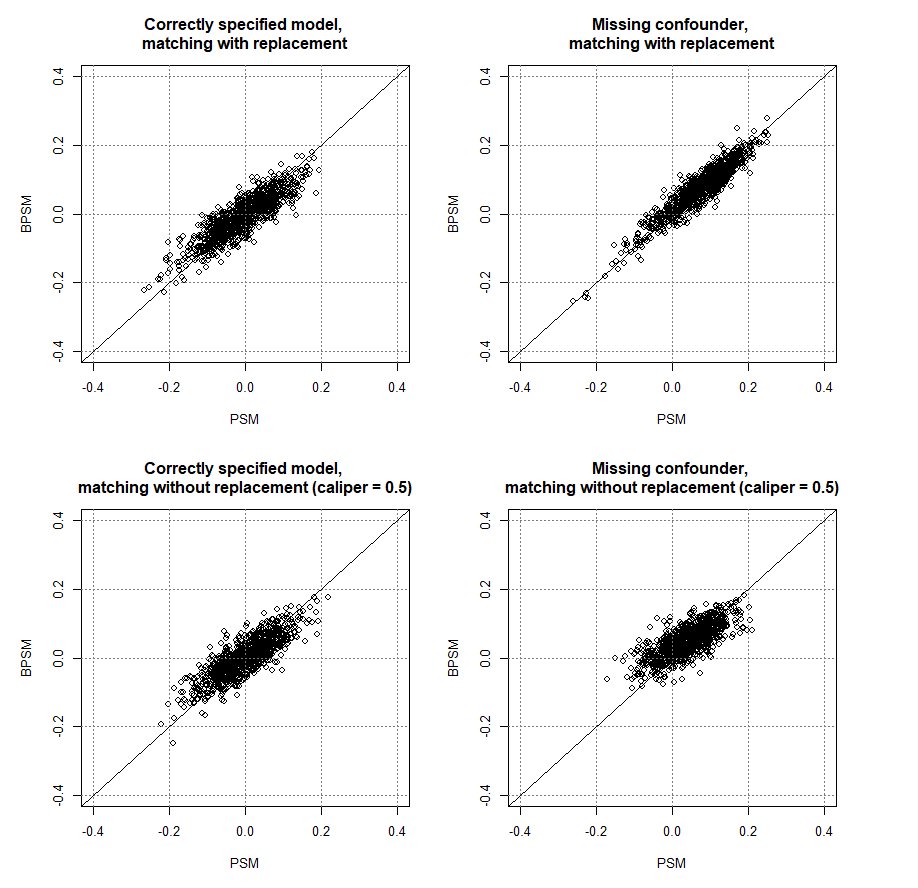}
    \centering{\parbox{5in}{\footnotesize{\vspace{0.5cm} Note: Each point corresponds to a different simulated dataset. Results correspond to a simulation study conducted under the assumption of small sample size ($N = 500$) and small treatment effect ($\beta = .25$). Number of simulations: 1,000.}}}
\label{fig:compbias}
\end{figure}


\begin{thebibliography}{99}

\bibitem[Abadie and Imbens(2008)]{abadie08} Abadie, Alberto and Guido W. Imbens. 2008. On the Failure of the Bootstrap for Matching Estimators. \emph{Econometrics} 76(6): 1537-57.

\bibitem[An(2010)]{an10} An, Weihua. 2010. Bayesian Propensity Score Estimators: Incorporating Uncertainties in Propensity Scores into Causal Inference. \emph{Sociological Methodology} 40(1): 151-189.

\bibitem[Albertus and Kaplan(2013)]{albertus13} Albertus, Michael and Oliver Kaplan. 2013. Land Reform as a Counterinsurgency Policy : Evidence from Colombia. \emph{Journal of Conflict Resolution} 57(2): 198-231.

\bibitem[Albertus and Kaplan(2013)]{albertus13r} Albertus, Michael and Oliver Kaplan. 2013b. Replication Data for ``Land Reform as a Counterinsurgency Policy: The Case of Colombia.'' Empirical Studies of Conflict Project (ESOC).

\bibitem[Cochran and Rubin(1973)]{cr73} Cochran, William G. and Donald B. Rubin. 1973. Controlling for Bias in Observational Studies: A Review. \emph{Sankhya: The Indian Jounal of Statistics: Series A} 35(4): 417-66.

\bibitem[Dehejia and Wahba(2002)]{dw02} Dehejia, Rajeev H. and Sadek Wahba. 2002. Propensity Score-Matching Methods for Nonexperimental Causal Studies. \emph{Review of Economics and Statistics} 84(1): 151-161.

\bibitem[Diamond and Sekhon(2013)]{ds13} Diamond, Alexis  and Jasjeet S. Sekhon. 2013. Genetic Matching for Estimating Causal Effects: A General Multivariate Matching Method for Achieving Balance in Observational Studies. \emph{Review of Economics and Statistics} 95(3): 932-45.

\bibitem[Heckman, Ichimura, and Todd(1997)]{hit97}  Heckman, James J., Hidehiko Ichimura, and Petra E. Todd. 1997. Matching As An Econometric Evaluation Estimator: Evidence from Evaluating a Job Training Programme. \emph{Review of Economic Studies} 64 (4): 605-54.

\bibitem[Ho et al.(2007)]{ho07} Ho, Daniel E., Kosuke Imai, Gary King, and Elizabeth A. Stuart. 2007. Matching as nonparametric preprocessing for reducing model dependence in parametric causal inference. \emph{Political Analysis} 15 199-236.

\bibitem[Iacus, King, and Porro(2012)]{ikp12} Iacus, Stefano M., Gary King, and Giuseppe Porro. 2012. Causal Inference without Balance Checking: Coarsened Exact Matching. \emph{Political Analysis} 20(1): 1-24.

\bibitem[Imai and Dyk(2004)]{id04} Imai, Kosuke and David A van Dyk. 2004. Causal Inference With General Treatment Regimes: Generalizing the Propensity Score. \emph{Journal of the American Statistical Association} 99(467): 854-66.

\bibitem[Imai and Ratkovic(2012)]{ir12} Imai, Kosuke and Marc Ratkovic. 2012. Covariate Balancing Propensity Score. \emph{Journal of the Royal Statistical Society: Series B (Statistical Methodology)} 76(1) 243-263

\bibitem[Jackman(2000)]{jackman00} Jackman, Simon. 2000. Estimation and Inference Are Missing Data Problems: Unifying Social Science Statistics via Bayesian Simulation. \emph{Political Analysis} 8(4):307-32.

\bibitem[Kaplan and Chen(2012)]{kaplan12} Kaplan, David and Jianshen Chen. 2012. A Two-Step Bayesian Approach for Propensity Score Analysis: Simulations and a Case Study. \emph{Psychometrica} 77(3): 581-609.

\bibitem[King and Nielsen(2019)]{king16} King, Gary and Richard Nielsen.  2019.  ``Why Propensity Scores Should Not Be Used for Matching."  Forthcoming in \emph{Political Analysis}, {\url doi:10.1017/pan.2019.11}.

\bibitem[McCandless et al.(2009)]{mccandless09} McCandless, Lawrence C., Paul Gustafson, and Peter C. Austin. 2009. Bayesian Propensity Score Analysis for Observational Data. \emph{Statistics in Medicine} 28(1): 94-112.

\bibitem[Rosenbaum(1999)]{r99} Rosenbaum, Paul R. 1999. Propensity Score. In Armitage P, Colton T, eds., \emph{Encyclopedia of Biostatistics.} New York, NY: John Wiley, 3551-5.

\bibitem[Rosenbaum and Rubin(1983)]{rr83} Rosenbaum, Paul R. and Donald B. Rubin. 1983. The Central Role of the Propensity Score in Observational Studies for Causal Effects. \emph{Biometrika} 70:41-55.

\bibitem[Rosenbaum and Rubin(1985)]{rr85} Rosenbaum, Paul R. and Donald B. Rubin. Constructing a Control Group Using Multivariate Matched Sampling Methods that Incorporate the Propensity Score.  \emph{American Statistician} 39:33-8.

\bibitem[Rubin(1973)]{r73} Donald B. Rubin. 1973. The Use of Matched Sampling and Regression Adjustment to Remove Bias in Observational Studies. \emph{Biometrics} 29(1): 185-203.

\bibitem[Rubin(1979)]{r79} Donald B. Rubin. 1979. Using Multivariate Matched Sampling and Regression Adjustment to Control Bias in Observational Studies. \emph{Journal of the American Statistical Association} 74(366): 318-28.

\bibitem[Stuart(2010)]{s10} Stuart, Elizabeth A. 2010. Matching Methods for Causal Inference: A Review and a Look Forward. \emph{Statistical Science} 25(1):1-21

\bibitem[Tu and Zhou(2002)]{tu02} Tu, Wanzhu and Xiao-Hua Zhou. 2002. A Bootstrap Confidence Interval Procedure for the Treatment Effect Using Propensity Score Subclassification. \emph{Health Services \& Outcomes Research Methodology} 3: 135-147.

\bibitem[Zigler and Dominici(2014)]{zd14} Zigler, Corwin Matthew and Francesca Dominici. 2014. Uncertainty in Propensity Score Estimation: Bayesian Methods for Variable Selection and Model-Averaged Causal Effects. \emph{Journal of the American Statistical Association} 109(505): 95-107.

\end{thebibliography}
\end{document}